\documentclass[10pt,twocolumn]{revtex4}
\usepackage[utf8]{inputenc}
\usepackage{amsmath}
\usepackage{amssymb}
\usepackage{graphicx}
\usepackage{natbib}
\DeclareMathOperator{\sech}{sech}
\DeclareMathOperator{\re}{Re}

\begin{document}
\title{Pulse growth dynamics in laser mode locking}
\author{Mark Popov and Omri Gat}
\affiliation{ Racah Institute of Physics, The Hebrew University, Jerusalem 9190401, Israel}

\begin{abstract}
We analyze theoretically and numerically the nonlinear process of pulse formation in mode locked lasers starting from a perturbation of a continuous wave. Focusing on weak to moderate dispersion systems, we show that pulse growth is initially slow, dominated by a cascade of energy from low to high axial modes, followed by fast strongly nonlinear growth and finally relaxation to the stable pulse waveform. The pulse grows initially by condensing a fixed amount of energy into a decreasing time interval, with peak power growing toward a finite-time singularity that is checked when the gain bandwidth is saturated by the pulse. 
\end{abstract}
\maketitle
\paragraph*{Introduction}
Passive mode locking is essential for the creation of ultrashort pulses, and as such has long been the subject of intense study \cite{Haus:2000va,Kutz:2006gh}. The complex evolution of the optical waveform in passive mode locked lasers (PML) is an extended nonlinear dynamical systems. While the stationary pulse states of PML are well-understood \cite{Grelu:2012gc,Turitsyn:2016ff}, a systematic theory of the rich variety of transient and unsteady states of PML is still lacking.

Here we focus on a transient process of particular importance---the buildup of an ultrashort pulse from an initial weak narrow band quasi-continuous wave (cw). The pulse growth dynamics has been studied in several papers, starting in the 1990's \cite{Krausz:1991tq,Anonymous:RnsgHeds,Brabec:1991vd,KUO:1992vc,SARUKURA:1992tm,CERULLO:1994tg,PU:1995td,Sutter:1998wf,Sun:2001uj}, where the main goal was to study the question of self-starting; this question was later answered by a statistical mechanics analysis of the laser, where mode locking is a thermodynamic phase transition \cite{Gordon:2002cg,Gat:2004co,Vodonos:2005tg,Gat:2005kx,Gordon:2006uz}.

The early studies focused on the early stages of the growth process, and experimental data on the later evolution of the waveform has been limited.
More recently there has been a resurgence of interest in the subject with the advent of experimental techniques allowing for real-time tracking of the evolving optical waveform \cite{Li:2010tq,Jasik:2013kn,Herink:2016ku,Ryczkowski:2017ur}. 

On the other hand, the theoretical understanding of the pulse growth process is rudimentary: From a dynamical systems perspective \cite{Wang:2014ki}, the waveform trajectory is a heteroclinic orbit connecting an unstable fixed point, continuous wave (cw), and a stable pulse fixed point. The fixed points, that are stationary solutions of the governing equations, are well studied, and so is the linearized dynamics very close to them. However the bulk of the evolution dynamics is nonlinear, and exhibits considerable complexity \cite{Li:2010tq,Jasik:2013kn,Herink:2016ku,Ryczkowski:2017ur}. 

We tackle this problem numerically and theoretically, in an approach based the Haus mode locking model~\cite{Haus:2000va}. 
The first key conclusion is that the nature of pulse growth depends critically on the strength of the dispersive effects---group velocity dispersion and Kerr nonlinearity---relative to the gain and loss. When the dispersive effects are weak or comparable to the gain and loss terms, the peak power of the pulse grows monotonically, and the pulse width decreases monotonically, as the waveform evolves from cw to pulse, and the growth dynamics is reproducible; when dispersive effects are strong on the other hand, the waveform evolves in an oscillatory and chaotic manner, and the pulse generation is a stochastic event. The strong dispersion regime is typical for ultrashort pulse lasers \cite{Herink:2016ku}.

Here we study the simpler weak-to-moderate dispersion regime, which is relevant to picosecond pulse lasers. We show that although the final stages of pulse formation occur on a fast nonlinear time scale $\tau_{NL}$ of order $\mu$s, it is preceded by a slow growth process that is one-to-two orders of magnitudes slower. Moreover, whereas the final fast growth is strongly nonlinear, the initial slow dynamics is governed by a small parameter, the ratio of the spectral width to the final bandwidth.

We develop a weakly nonlinear theory for the slow growth process, showing that pulse buildup progresses through a cascade of power from lower to higher axial modes, driven by the absorptive and Kerr nonlinearities. We derive explicit recursive expressions for the mode amplitudes as series of growing exponentials, whose radius of convergence marks the transition to strongly nonlinear growth.

\begin{figure*}[tb]
\includegraphics[width=7.5cm]{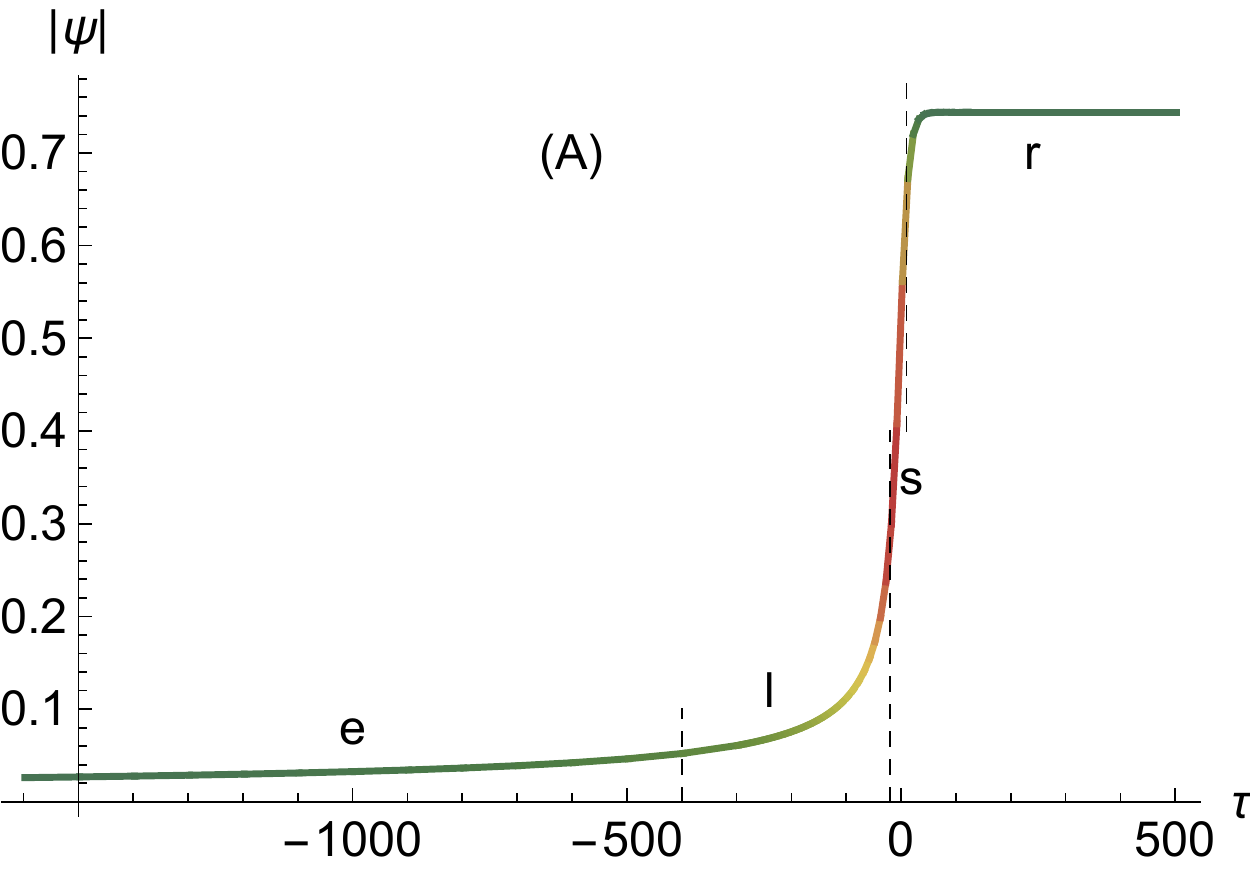}\qquad\includegraphics[width=7.5cm]{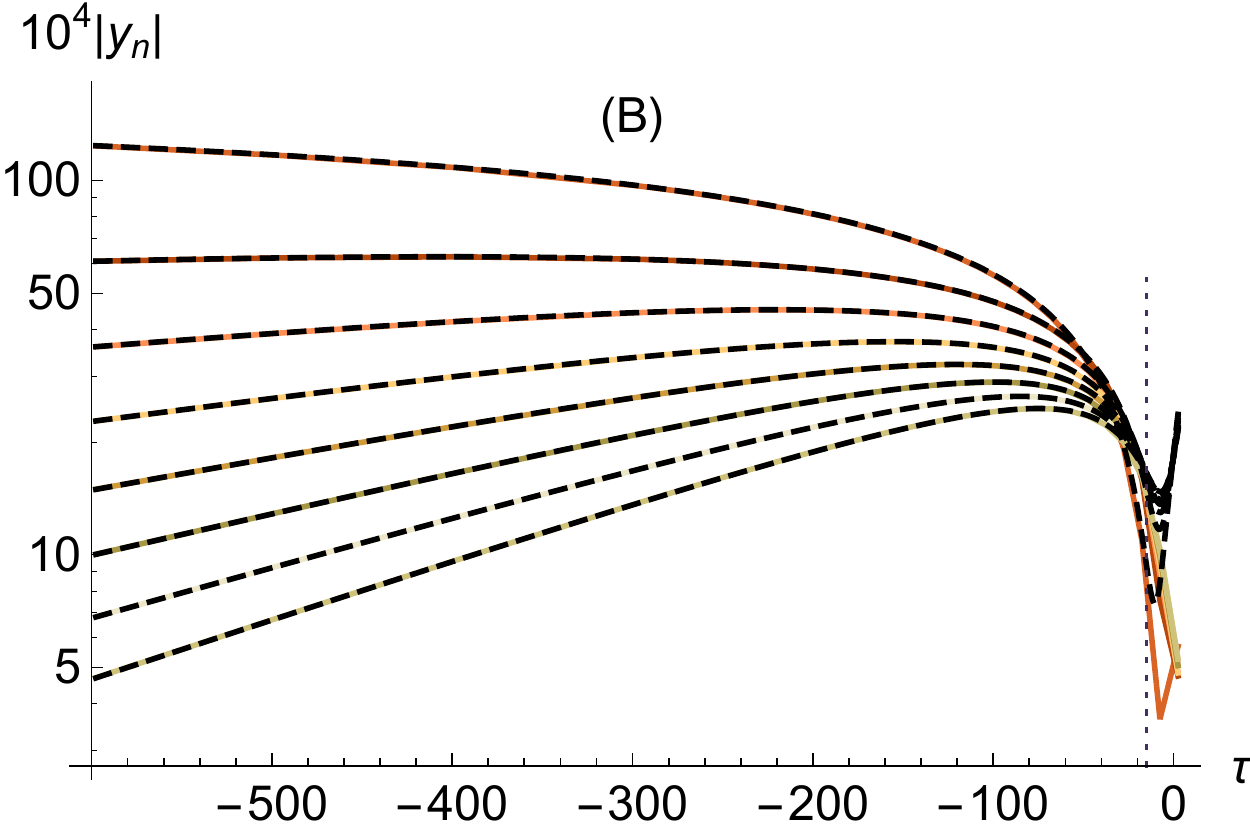}\\\includegraphics[width=7.5cm]{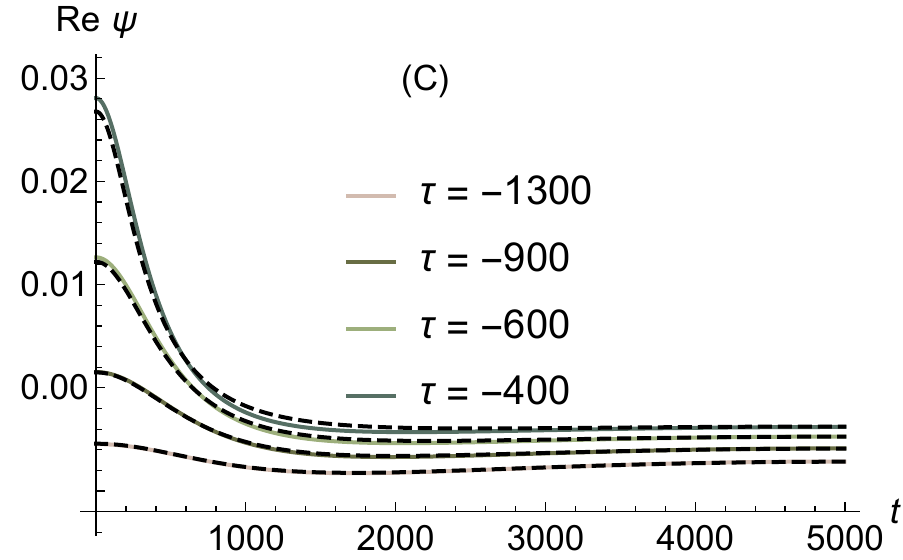}\quad\includegraphics[width=7.5cm]{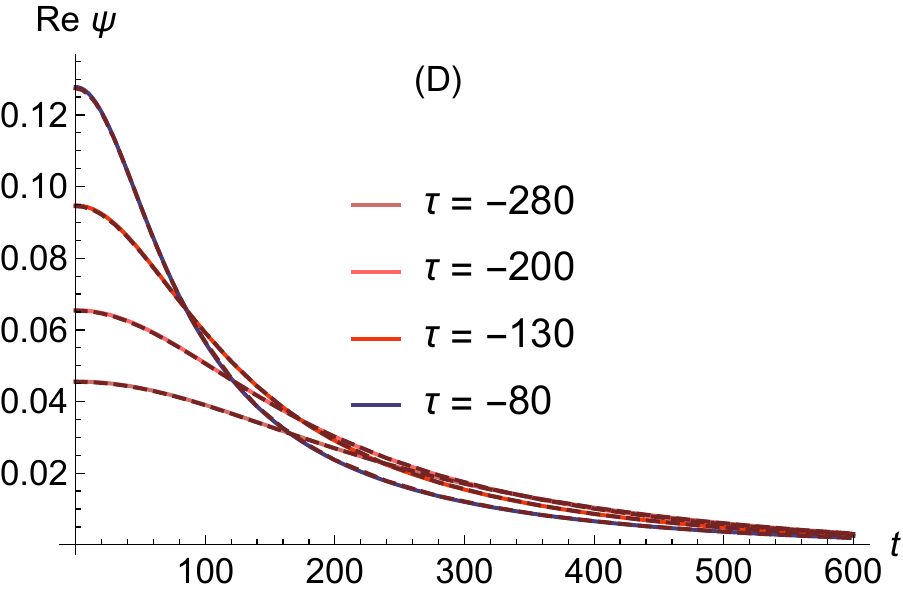}
\caption{\label{fig:f} Simulation results (solid curves) and corresponding theoretical calculation (dashed curves) for the pulse formation process with normalized dispersion $\beta=-1.5$, Kerr $\gamma=0.6$, gain parameter $k=2.7$, and chirp exponent $c=1.58$\,. (A) The numerically calculated peak amplitude $|\psi(0,\tau)|$ as a function of the slow time $\tau$. The growth stages marked on the graph are: \textbf{e}arly weakly nonlinear, \textbf{l}ate weakly nonlinear, \textbf{s}trongly nonlinear, and \textbf{r}elaxation. The origin of $\tau$ is defined as the time where the pulse amplitude reaches one half of its final value. (B) The eight leading Fourier amplitudes $|y_n|$, $n=0,1,\ldots,7$ ordered from top to bottom on the left, as a function of slow time $\tau$ ($y_0$ includes cw term.) (C) and (D) Snapshots of the real part of the waveform $\re\psi(t,\tau)$ as a function of fast time $t$ for several values of $\tau$ in the early (C) and late (D) weakly nonlinear growth stage. Note different scales between the two panels.
}
\end{figure*}

A principal result is that the slow process itself takes place in two stages. In the early stage the bandwidth of the pulse is so small that gain filtering and dispersion are negligible; it leads to a \emph{constant-power, form-invariant} pulse growth and narrowing toward a finite-time singularity.
It is averted by the gain filtering and dispersion, whose strength increases with the growing pulse bandwidth.
These theoretical predictions are compared with direct numerical simulations of the model, and shown to agree for several choices of model parameters in figures \ref{fig:f}--\ref{fig:l}. 

When the pulse peak power reaches 90\% of its final value, the pulse shape is close enough to its final form that its dynamics is well-approximated by relaxation of the normal modes of perturbation of the mode-locked state. In the cases studied here, the dominant, least stable mode, is a discrete mode, associated with amplitude perturbations \cite{CHEN:1994wk,Kapitula:2002ut,Wang:2014ki}. It leads to exponential convergence toward the final pulse shape.

\paragraph*{Laser model, steady states, and linear stability}
We model the laser with the Haus master mode locking equation
\begin{multline}\label{eq:haus}
\frac{\partial\psi}{\partial\tau}=\biggl(\frac{g_u}{1+P[\psi]/P_s}\Bigl(1-\frac{1}{\omega_g^2}\frac{\partial^2}{\partial t^2}\Bigr)\\-l-{i\beta_2}\frac{\partial^2}{\partial t^2}+(\gamma_s+i\gamma_k)\lvert \psi\rvert^2\biggr)\psi
\end{multline}
in which $\psi$ is the electric field envelope depending on the fast and slow time coordinates $t$, and $\tau$ (respectively), and $g_u$, $l$, $P_0$, $\omega_g$, $\beta_2$, $\gamma_s$ and $\gamma_k$ are the small signal gain and loss coefficients, gain saturation power, gain bandwidth, group velocity dispersion, fast absorber saturability, and Kerr coefficient, respectively.
The gain recovery time is assumed to be much longer than the round trip time $\tau_R$, so that the saturated gain depends only on the total power $P[\psi]=(1/\tau_R)\int_0^{\tau_R}\lvert \psi\rvert^2dt$. 


The simplest steady state of Eq.\ \eqref{eq:haus} is the dark state $\psi\equiv0$, which is subject to the lasing instability, where the amplitude of an axial mode with frequency $\omega$ grows with rate $g_u(1-\omega^2/\omega_g^2)-l$ (when this rate is positive) and a random phase. The growth of the lasing instability eventually leads to the cw state. In the Haus model \eqref{eq:haus} the cw state is a waveform with a uniform amplitude $a_c$.

We now make the simplifying assumption, which usually holds in experiments, that the gain is deeply saturated, $P\gg P_s$,  and then choose $1/\omega_g$, $1/l$, and $(l/\gamma_s)^{1/2}$ as units of $t$, $\tau$, and $\psi$, respectively, so that the equation depends on four dimensionless combinations, $\beta=-\beta_2\omega_g^2/l$, $\gamma=\gamma_k/\gamma_s$, $k=g_uP_s\gamma_s\tau_R\omega_g/l^2$, and $r=\tau_R\omega_g$, becoming
\begin{equation}\label{eq:haus-dl}
\frac{\partial\psi}{\partial\tau}=\left(\frac{k}{\lVert \psi\rVert^2}(1+\frac{\partial^2}{\partial t^2})-1+{i\beta}\frac{\partial^2}{\partial t^2}+(1+i\gamma)\lvert \psi\rvert^2\right)\psi
\end{equation}
where $\lVert \psi\rVert^2=\int_0^{r}\lvert \psi\rvert^2dt$ (note that $\beta>0$ corresponds to anomalous dispersion.) $r$ is the cavity roundtrip time in the natural choice of units, and as such does not appear explicitly in \eqref{eq:haus-dl}; it effectively measures the number of active modes, so we let $r\gg1$, leaving three order-1 parameters.

In this parametrization the cw solution is 
\begin{equation}\label{eq:ac}
\psi_c(t,\tau)=a_ce^{i\Omega_c\tau}\ ,\quad\frac{k}{a_c^2r}=1-a_c^2\ ,\quad \Omega_c=\gamma a_c^2\ ,
\end{equation}
which implies that $a_c\sim\sqrt{k/r}\ll1$; here $\Omega_c$ is the cw nonlinear frequency shift.

To study perturbation to cw we write
\begin{equation}
\psi(t,\tau)=e^{i\Omega_c\tau}\sum_n (y_n(\tau)+\delta_{n,0}a_c)e^{i\omega_nt}\ ,\quad \omega_n=2\pi n/r\ ,
\end{equation}
where $\delta$ is the usual Kronecker delta; an infinitesimal perturbation to the cw evolves according to 
\begin{align}\label{eq:haus-linearized0}
y_0'(\tau)&=(-1+(1+i\gamma) a_c^2)(y_0+y_0^*)\ ,\\ 
y_n'(\tau)&=\bigl((1+i\gamma)a_c^2-(1+i\beta)\omega_n^2\bigr)y_n\nonumber\\&\qquad\qquad\qquad\qquad+(1+i\gamma)a_c^2y_n^*\ ,\quad n\ne0\ ,\label{eq:haus-linearized}
\end{align}
where $\omega_n=2\pi/r$. It follows that $y_0$ decays, while the growth rate of $y_n$, $n\ne0$,
\begin{equation}
\lambda_n=a_c^2-\omega_n^2+\sqrt{a_c^4(1+\gamma^2)-(\gamma a_c^2-\beta\omega_n^2)^2}\ ,
\end{equation}
is positive for sufficiently low modes since $\omega_n\ll r^{-1/2}\sim a_c$, making the cw modulationally unstable. We will assume that $\beta\gamma<1$ and then $\lambda_1=\lambda_{-1}$ are the largest growth rates.

We are interested in parameter values for which Eq.~\eqref{eq:haus-dl} has stable pulse steady-state solutions; these pulses are chirped-soliton shaped, 
\begin{equation}
\psi_p(t,\tau)=a_p\sech(bt)^{1+ic}e^{i\Omega_p\tau}\ .
\end{equation}
 The pulse amplitude $a_p$, width $1/b$, chirp $c$ and nonlinear frequency shift $\Omega_p$ are determined by the physical parameters \cite{Haus:2000va}. For fixed values of $\beta,\gamma$, and small enough $k$, there are usually two pulse solutions, of which the one with lower energy is stable \cite{Wang:2014ki}.

\paragraph*{Numerical calculations.}
We studied the formation of pulses by numerically solving Eq.\ (\ref{eq:haus-dl})  for several values of the (normalized) dispersion $\beta$ and Kerr coefficient $\gamma$, yielding pulses with negative, zero, and positive chirp. For each choice of $\beta$ and $\gamma$ there is maximal value of the gain parameter $k$ that admits stable pulses, and the results shown here are for $k$ close to this maximum. We checked that the dynamics is weakly sensitive to changes in $k$.

In each of the simulations, the roundtrip time was $r=10^4$ natural units, which corresponds to physical repetition rate of around 100 MHz, and the initial waveform was the cw solution perturbed along the most unstable mode with an amplitude of $10^{-8}$ natural units, corresponding to fluctuations of the order $10^{-7}$ W in the mode. The wave equation was solved with two methods, 4th/5th order Runge-Kutta with variable grid, and pseudo-spectral split step, with compatible results. Results reported here were obtained with the first method.

The simulations results are presented in figures \ref{fig:f}--\ref{fig:l}, with panels A showing the growth of the peak amplitude of the pulse, panels B the evolution of the amplitudes of the higher Fourier modes, and panels C and D the evolution of the real part of the waveform. Panels B, C, and D compare the simulation results with weakly nonlinear theory, showing very good agreement.

\begin{figure*}[tb]
\includegraphics[width=7.5cm]{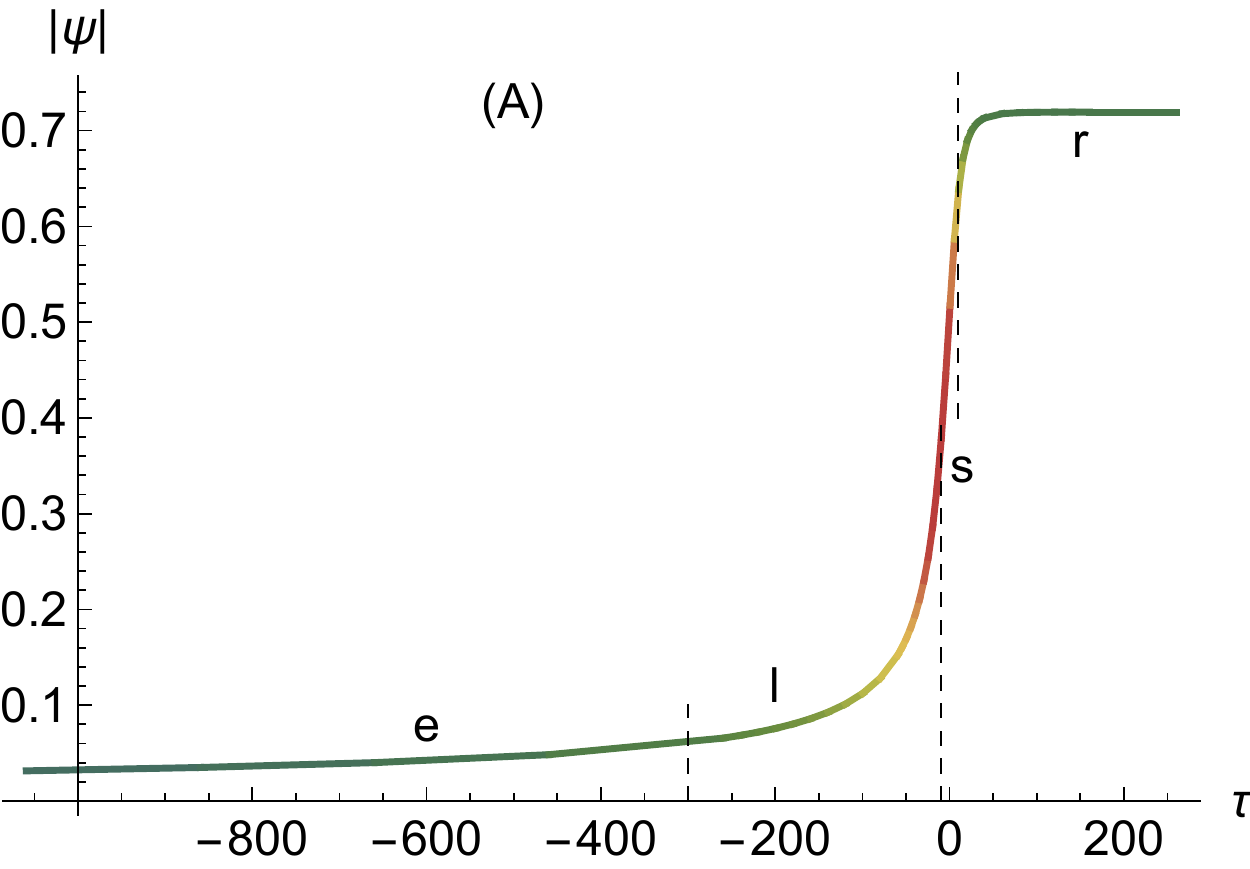}\quad\includegraphics[width=7.5cm]{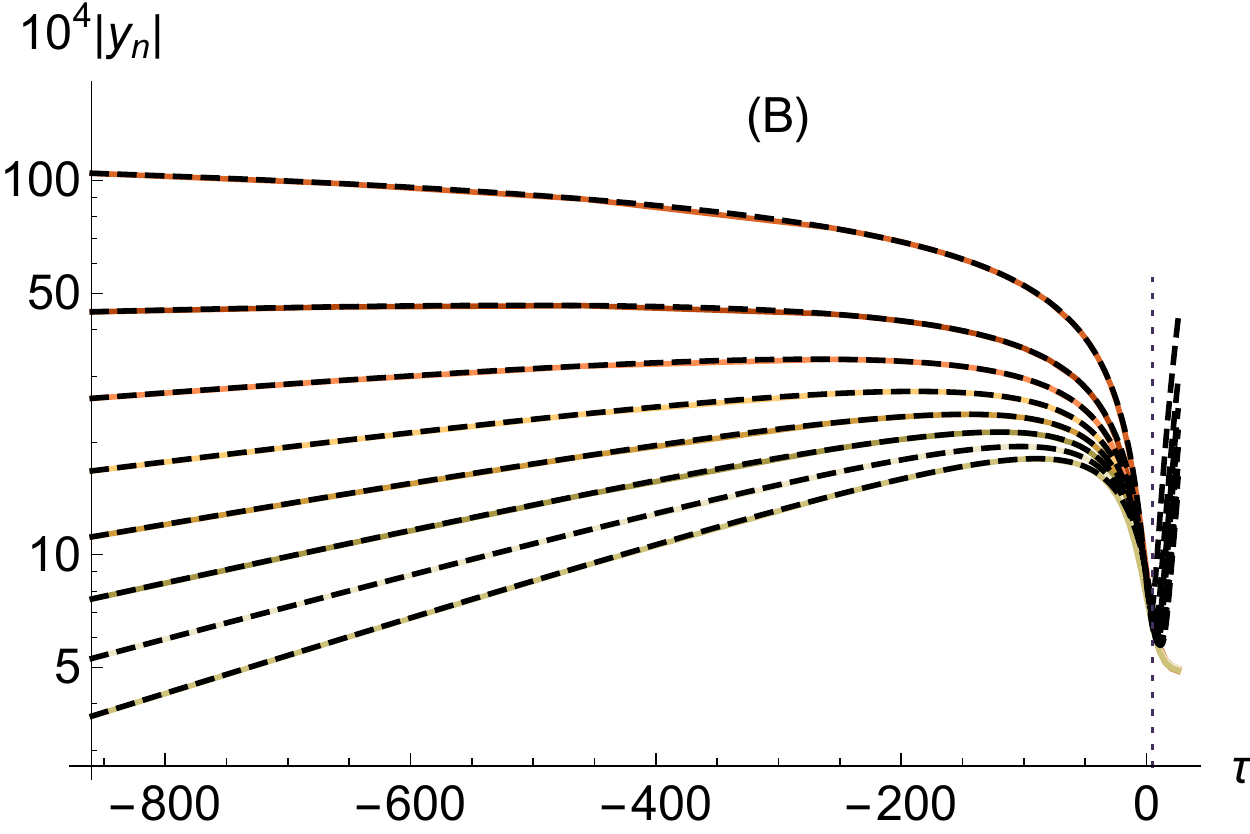}\\\includegraphics[width=7.5cm]{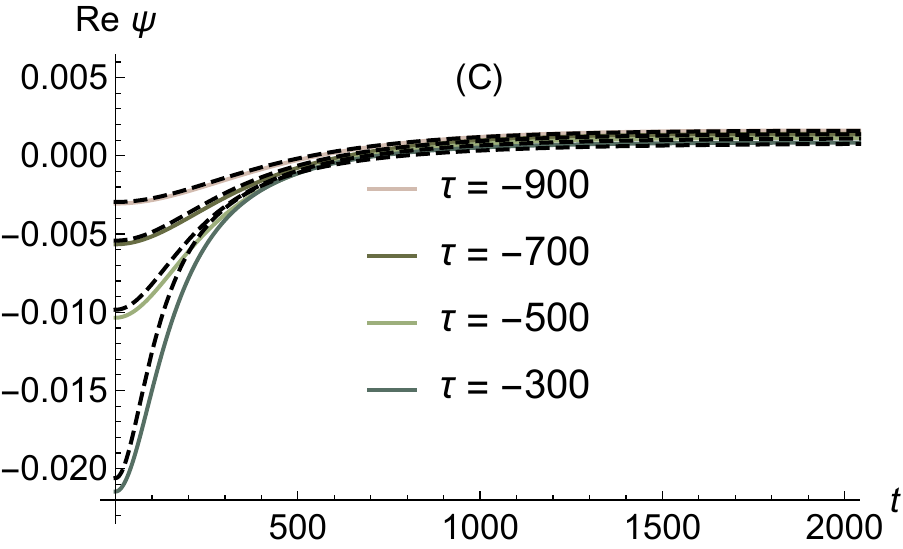}\qquad\includegraphics[width=7.5cm]{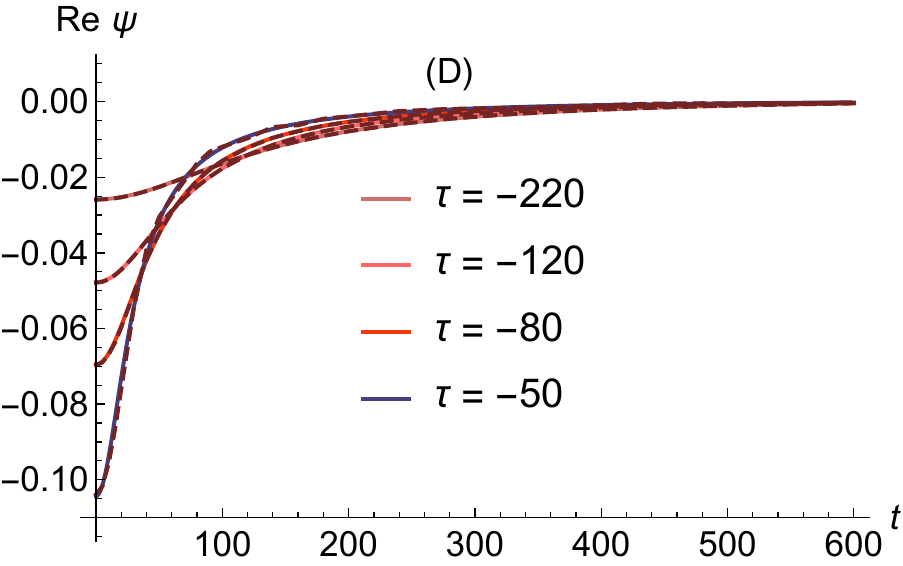}
\caption{\label{fig:l} Simulations and theory for pulse growth with $\beta=1.0$, $\gamma=0.2$, $k=1.72$, $c=-0.53$. See Fig.\ \ref{fig:f} for panel descriptions}
\end{figure*}
\paragraph*{Mode energy cascade in weakly nonlinear pulse growth}
For the weakly nonlinear analysis we assume that the $y_n$s are small but not infinitesimal, write Eq.~\eqref{eq:haus-dl} in terms of mode amplitudes\begin{multline}\label{eq:haus-modes}
y_n'(\tau)+i\Omega_c y_n(\tau)=\left(g(1-{\omega_n^2})-1-{i\beta\omega_n^2}\right)y_n\\+(1+i\gamma)\sum_{k-l+m=n}y_ky_l^*y_m\ ,
\end{multline}
and Taylor expand the gain
\begin{multline}\label{eq:gexp}
g=\frac{a_c^2(1-a_c^2)}{|a_c+y_0|^2+\sum_{k\ne0}\lvert y_k \rvert^2}=
(1-a_c^2)\\\times\Bigl(1+\sum_j(-\frac{2\re y_0}{a_c}-\frac{| y_0|^2+\sum_{k\ne0}|y_k|^2}{a_c^2})^j\Bigr)\ ,
\end{multline}
As the pulse waveform is an even function, we will assume for simplicity that $\psi$ is an even function of $t$ initially, and therefore for all $\tau$, so that $y_{-n}(\tau)=y_n(\tau)$.

Since $y_1$ is the fastest growing mode, there is time interval where it dominates all higher modes, while still being a small perturbation of the cw amplitude, 
\begin{equation}\label{eq:wnlg}
a_c\gg |y_{1}(\tau)|\gg|y_n(\tau)|,\quad  n\ge2\ .
\end{equation}  
For such $\tau$ values $y_1$ grows according to the linearized dynamics Eq.\ \eqref{eq:haus-linearized}, so that
\begin{equation}\label{eq:y1}
y_{1}(z)=\alpha_{1}(1+i\gamma)e^{\lambda\tau}\ ,\qquad \lambda=2a_c^2\ ,\quad \alpha_1\text{ real },
\end{equation}
(neglecting $\omega_1^2$ with respect to $a_c^2$.)

Consider now \eqref{eq:haus-modes} with $n=2$. When \eqref{eq:wnlg} holds, the largest terms are those proportional to $y_2$, and the terms in the triple sum with $0\le k,-l,m\le 1$. Neglecting all other nonlinearities, and using $\omega_2\ll a_c$, the equation for $y_2$ becomes
\begin{multline}
y_2'(z)
=(1+i\gamma)a_c^2(y_2+y_2^*)\\+(1+i\gamma)\bigl(2(1+\gamma^2)+(1+i\gamma)^2)a_c\alpha_1^2e^{2\lambda\tau}\ ,
\end{multline}
using \eqref{eq:y1}. Since the nonlinear growth rate in this equation, $2\lambda$ is faster than the linear growth rate, $\lambda_2$, of $y_2$, it eventually dominates the dynamics, so that for large $\tau$
\begin{equation}
 y_2=\frac{\alpha_1^2}{2a_c}(1+i\gamma)(3+i\gamma)e^{2\lambda\tau}\equiv\alpha_2(1+i\gamma)e^{2\lambda\tau}
\end{equation}

Proceeding in the same fashion for higher $n$ we find
\begin{align}\label{eq:alphan}
y_n(z)&\sim \alpha_n(1+i\gamma)e^{n\lambda\tau}\\\alpha_n&=\frac{(3+i\gamma)\cdots(2n-1+i\gamma)}{n!}\frac{\alpha_1^n}{a_c^{n-1}}\ ,\quad n\ge0\ ,
\end{align}
with the convention $\alpha_0=a_c$. 

Our next goal is to calculate the waveform $\psi(t,\tau)$. It is not sufficient for this purpose to sum the leading term for each mode since subleading terms in $y_n$ can be as large or larger than the leading term in $y_m$, $m>n$. We therefore look for solutions of the form
\begin{equation}\label{eq:alphannu}
y_n(\tau)=\sum_{\nu=0}^\infty\alpha_{n,\nu}(1+i\gamma)e^{(n+2\nu)\lambda\tau}\ ,\end{equation}
where $\alpha_{n,0}=\alpha_n$ of Eq.\ \eqref{eq:alphan}. Plugging this expansion in \eqref{eq:haus-modes} yields a system of equations for the coefficients in which $\alpha_{n,\nu}$ is determined by $\alpha_{n+1,\nu-1},\ldots,\alpha_{n+\nu,0}$, which can be solved recursively. 
The results of this calculation are in excellent agreement with direct simulations, as shown in panel B of figures \ref{fig:f}--\ref{fig:l}.
\begin{figure*}[tb]
\includegraphics[width=7.5cm]{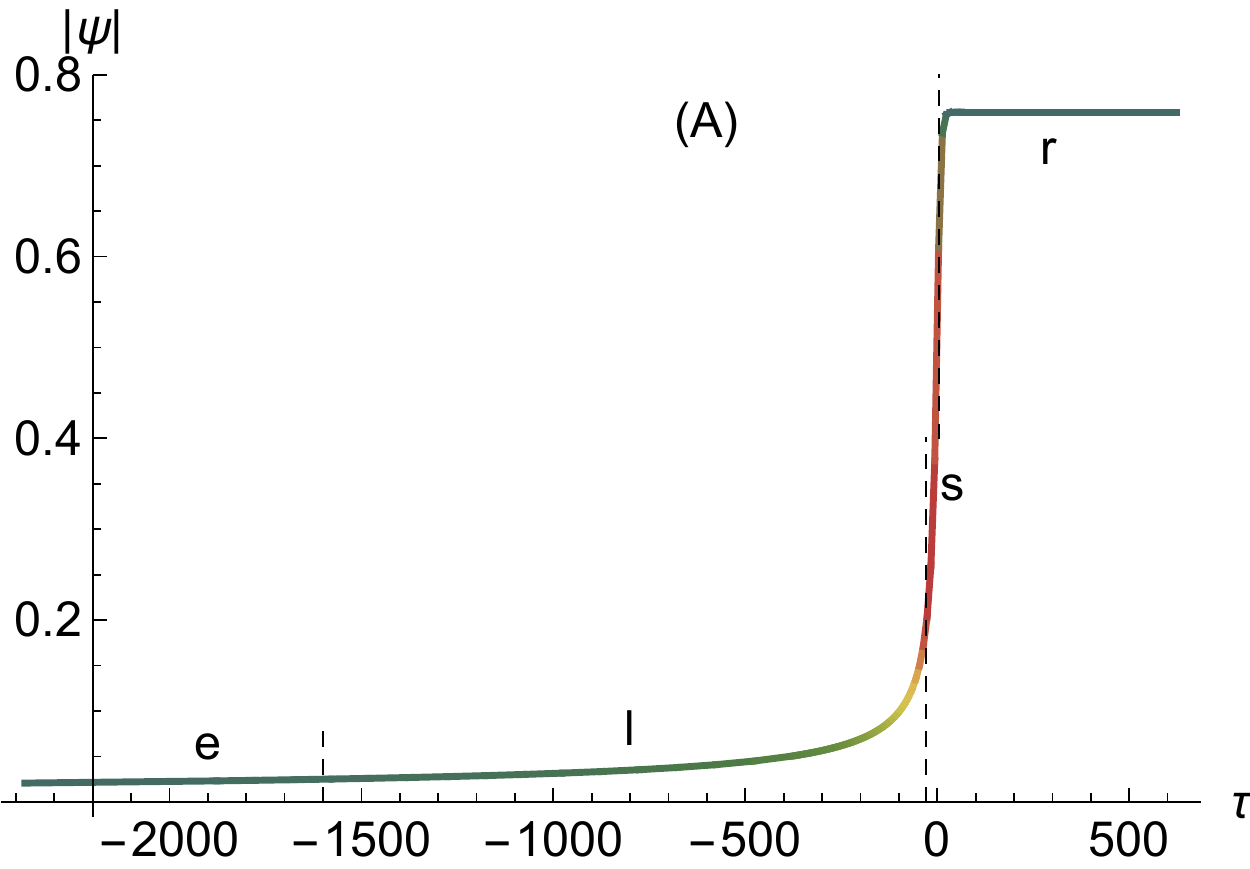}\qquad\includegraphics[width=7.5cm]{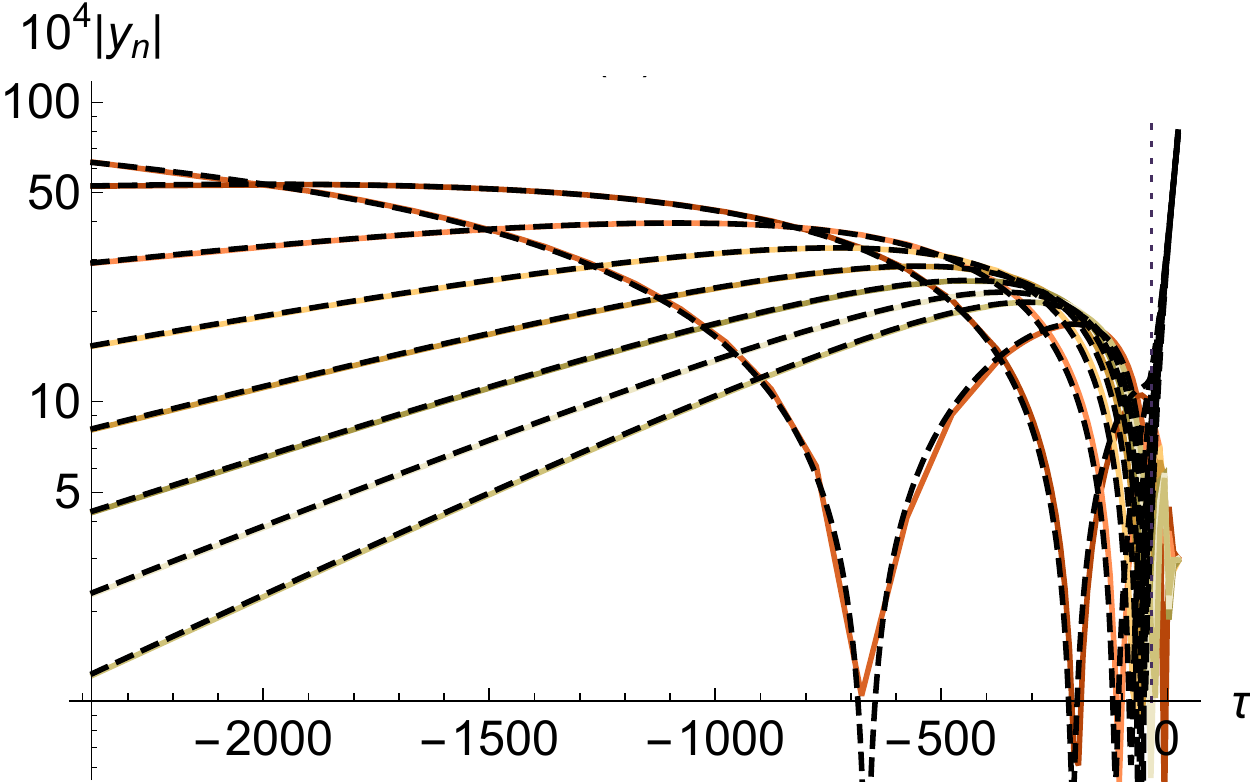}\\\includegraphics[width=7.5cm]{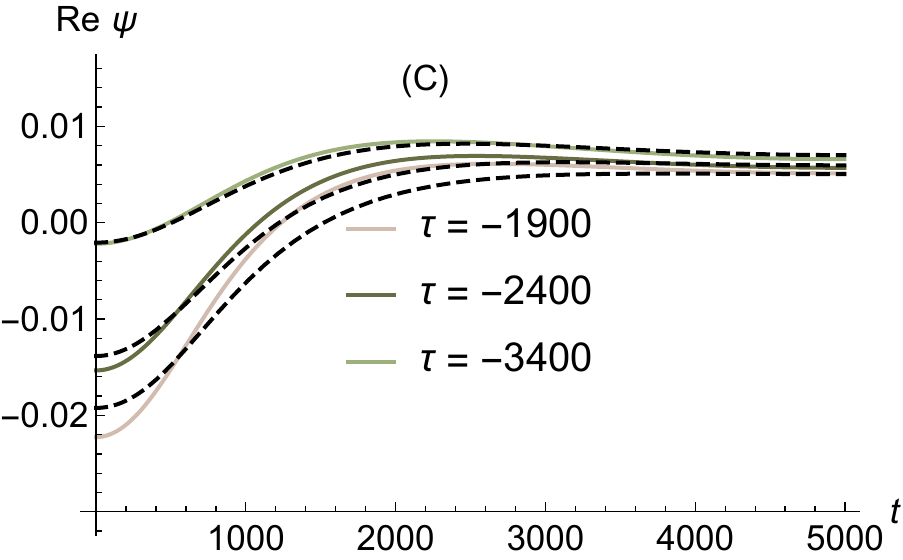}\quad\includegraphics[width=7.5cm]{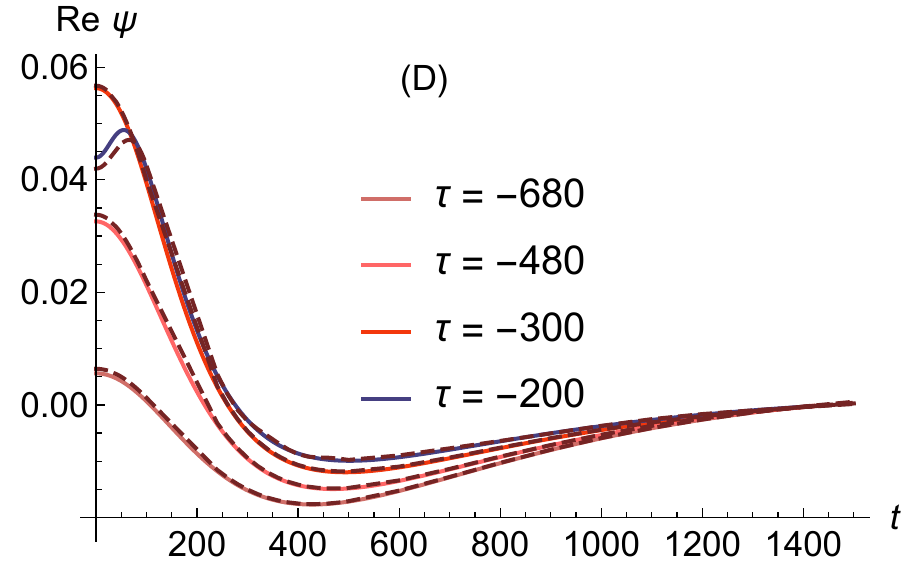}
\caption{\label{fig:l} Simulations and theory for pulse growth with $\beta=0.5$, $\gamma=2.0$, $k=1.08$, $c=0.34$. See Fig.\ \ref{fig:f} for panel descriptions.}
\end{figure*}

In the early stage of pulse growth, the bandwidth is small, and terms proportional to $\omega_n^2$ are negligible in~\eqref{eq:haus-modes}. Although we could not find a closed form expression for the coefficients $\alpha_{n,\nu}$, we verified order by order that, when gain bandwidth and dispersion are neglected, the coefficients $\alpha_{n,\nu}$ are such that  \eqref{eq:alphan} \emph{still holds}, up to an overall multiplicative factor, and moreover, the overall scale is such that the mean power remains \emph{constant},
\begin{equation}\label{eq:meanpower}
\frac{1}{r}\int dt|\psi(t,\tau)|^2dt=\sum_n|y_n(\tau)|^2=a_c^2\ .
\end{equation}
Summing the Fourier series defined by Eq.\ \eqref{eq:alphannu} then gives the early stage waveform explicitly
\begin{align}\label{eq:early}
&\psi_e(t,\tau)= a_w(t)\Bigl(\bigl(1-\frac{2\alpha_1(1+i\gamma)}{a_c}e^{i\omega t+\lambda\tau}\bigr)^{-(1+i\gamma)/2}\nonumber\\&\qquad+\bigl(1-\frac{2\alpha_1(1+i\gamma)}{a_c}e^{-i\omega t+\lambda\tau}\bigr)^{-(1+i\gamma)/2}-1\Bigr)\ ,
\end{align}
where 
\begin{equation}
|a_w|=a_c(2F(\frac12(1-i\gamma),\frac12(1-i\gamma);1;4(\alpha_1/a_c)e^{2\lambda\tau}-1)
\end{equation}
is fixed by \eqref{eq:meanpower} (here $F$ is the standard Gauss hypergeometric function.) This result is verified by comparison with simulations, in panel C of figures \ref{fig:f}--\ref{fig:l}.

This where a fixed amount energy is squeezed in a decreasing time interval becomes singular at $\tau_\text{sing}$, defined by $2\alpha_1|1+i\gamma|e^{\lambda\tau_\text{sing}}=a_c$. As the waveform approaches singularity, however, the bandwidth diverges limitations of gain bandwidth and  dispersion inevitably smoothen the pulse shape. The expansion \eqref{eq:alphannu} is still valid at this point, the late stage of the weakly nonlinear growth, as shown in panel D of figures  \ref{fig:f}--\ref{fig:l}, but the pulse no longer has a simple invariant form. Interestingly, although gain filtering and dispersion check the growth of the pulse peak power, they cause the pulse energy to grow by an order-1 factor toward the final pulse energy.

\paragraph*{Strongly nonlinear growth and relaxation.}
The expansion \eqref{eq:alphannu} has a finite radius of convergence, whose boundary is evident in the right edge of panel B of figures~\ref{fig:f}--\ref{fig:l}. Beyond this point there is no small parameter, and the pulse dynamics becomes independent of the initial condition. This strongly nonlinear growth is much faster than the preceding dynamics, with the  amplitude growing close to its final value in approximately 50 natural time units. Further study of the strongly nonlinear growth process, and the relaxation to the final pulse waveform that follows it is beyond the scope of this work.

\paragraph*{Conclusions.}
Saturable absorbers are a very effective means to produce strongly nonlinear ultrashort pulses, but the initial pulse growth they produce by modulational instability is very weak because the initial intensity contrasts are small. The weak initial growth can be inhibited by noise or by chaos, and it is what makes self-starting of passive mode locking hard.

Here we study the growth of perturbations where effects of noise and chaos are negligible. Although this assumption is strictly valid only for picosecond lasers, we conjecture that the mode cascade buildup mechanism discovered here also works in ultrashort lasers, once the initial barrier to self starting has been crossed.

Since the linear growth rate is inversely proportional to the number of active modes, the duration of the growth process consists mainly of the time taken for the initial perturbation to grow to intensities that saturate the absorber. Nevertheless, the growth is not a linear process, except in its very early stage, when all modes amplitudes are comparable. Once the fastest growing mode dominates, it starts driving harmonics by four-wave mixing, making them grow much faster than the linear growth rate, so that the higher harmonics eventually catch up with the fundamental toward the creation of short pulse.

Moreover, during most of the weakly nonlinear growth each Fourier mode grows exponentially up to an overall constant factor. As a consequence, the pulse maintains a fixed functional form and total energy while narrowing and growing in amplitude.
 When the pulse amplitude reaches about 10\% of the final amplitude, higher order terms are no longer negligible, and the pulse is no longer form-invariant, and when the pulse amplitude further grows to about 30\% of its final value, the weakly nonlinear perturbation series ceases to converge altogether, and the much faster to strongly nonlinear growth begins. 
The last stage of pulse dynamics consists of relaxation toward the final stable shape and amplitude, and is characterized by a third timescale.

This work sheds light on the complex and sparsely studied process of pulse growth in nonlinear systems, with important implications to starting of mode locked lasers. It is likely that weakly nonlinear methods would facilitate a theory of the relaxation process. While the strongly nonlinear growth is by definition beyond the reach of perturbative analysis, it is possible that asymptotic approximation can be derived by resummation of the series approximation developed here.

\paragraph*{Acknowledgments}Nathan Kutz participated in early stages of this work, and we thank him for his contribution. We thank the US-Israel Binational Science Foundation for financial support

\bibliography{starting-growth}

\end{document}